\documentclass[prb,preprint,groupedaddress,showpacs]{revtex4}
\usepackage{graphicx} 
\begin{document} 
\bibliographystyle{apsrev}
 
\title{Spin-orbit effects on the Larmor dispersion relation in GaAs quantum wells}

\author{Francesc Malet} 
\affiliation{Departament ECM, Facultat de F\'{\i}sica, Universitat de Barcelona. 
Diagonal 647, 08028 Barcelona, Spain}

\author{Enrico Lipparini} 
\altaffiliation{Permanent address:
Dipartimento di Fisica, Universit\`a di Trento, and INFN,
38050 Povo, Trento, Italy}
\affiliation{Departament ECM, Facultat de F\'{\i}sica, Universitat de Barcelona. Diagonal 647,
08028 Barcelona, Spain} 

\author{Manuel Barranco} 
\affiliation{Departament ECM, Facultat de F\'{\i}sica, Universitat de Barcelona. Diagonal 647,
08028 Barcelona, Spain}

\author{Mart\'{\i} Pi} 
\affiliation{Departament ECM, Facultat de F\'{\i}sica, Universitat de Barcelona. Diagonal 647,
08028 Barcelona, Spain}

\date{\today}

\begin{abstract} 
We have studied the relevance of spin-orbit coupling to the dispersion 
relation of the Larmor resonance observed in inelastic light scattering 
and electron-spin resonance experiments on GaAs quantum wells.
We show that the spin-orbit interaction, here described by a sum of Dresselhaus
and Bychkov-Rashba terms, couples Zeeman and spin-density excitations.
We have evaluated its contribution to the spin splitting as a function of
the magnetic field $B$, and have found that in the small $B$ limit,
the spin-orbit interaction does not contribute to the spin splitting,
whereas at high magnetic fields it yields a $B$ independent contribution 
to the spin splitting given by $2(\lambda_R^2-\lambda_D^2)$, with
$\lambda_{R,D}$ being the intensity of the
Bychkov-Rashba and Dresselhaus spin-orbit terms.
\end{abstract}


\maketitle

\section{Introduction}

The study of spin-orbit (SO) effects in semiconductor nanostructures has 
been the object of many experimental and 
theoretical investigations in the last few years, see e.g. 
Refs. \onlinecite{Can99,Ric99,And99,Vos00,Mal00,Rac97, 
Vos01,Fol01,Hal01,Ale01,Val02,Val202,Val302,Kon05} and Refs. 
therein. It links the spin and the charge dynamics, hence 
opening the possibility of spin control by means of electric 
fields.\cite{Dat90,Kan98}

It has been recently shown\cite{Ton04} that the SO interaction affects 
the optical properties of GaAs quantum wells by
inducing a coupling between charge density and spin density 
excitations in the long wavelength limit. We extend 
here this study to the influence on the Larmor resonance 
of the combined effect of both 
Dresselhaus\cite{Dre55} and 
Bychkov-Rashba\cite{Ras84,Pik95} SO interactions,
and use our results to discuss some features 
of the spin modes disclosed by inelastic light scattering\cite{Dav97,Kan00}
and electron-spin resonance experiments.\cite{Ste82,Dob88}

Our approach is based on the solution of the equation of motion up to
second order in the SO intensity 
parameters.\cite{Ton04} This method has been also used to derive the Kohn
theorem,\cite{Pi04} and goes as follows. We write the
Schr\"odinger equation for a $N$-particle system as $H |n\rangle = E_n |n \rangle$, 
with $|0\rangle$ and $E_0$ being   the ground state (gs) 
and gs energy, respectively. If one can find an operator $O^+_n$ such that
$|n\rangle= O^+_n |0\rangle$, $O_n|0\rangle=0$, it is possible to cast the 
Schr\"odinger equation into an operator equation -the equation of motion- 
$[H,O^+_n] = \omega_n O^+_n$, where $ \omega_n= E_n -E_0$ is the excitation 
energy of the state $|n\rangle$. The solutions of this equation are used
to find the excitation energies of the system as well as
its excited states in terms of their creation 
operators.

This work is organized as follows. In Sec. II we apply the equation of motion 
method to the Larmor mode in the presence 
of a SO coupling. The results are used in Sec. III to discuss the spin 
modes in quantum wells, and are 
compared with the experimental results of Refs.~\onlinecite{Dav97,Dob88}.

\section{The equation of motion approach and the Larmor mode}

The operators describing the SO Rashba and Dresselhaus interactions are 
respectively given by
\begin{equation} 
\label{eq1}
H_R = \frac{\lambda_R}{\hbar}\sum_{j=1}^{N} \left[\, P_y\sigma_x-P_x\sigma_y\,\right]_j 
\end{equation} 
and 
\begin{equation} 
\label{eq2}
H_D = \frac{\lambda_D}{\hbar} \sum_{j=1}^{N} \left[\, P_x\sigma_x-P_y\sigma_y\,\right]_j~, 
\end{equation} 
where the $\sigma$'s are the Pauli matrices and 
${\bf P}=-i\hbar\nabla+\frac{e}{c}{\bf A}$ represents the 
canonical momentum in terms of the vector potential ${\bf A}$ which in the 
following we write in the Landau 
gauge, {\bf A}$= B (0,x,0)$, with {\bf B}=$ \nabla \times$ {\bf A} = $B \hat{\bf z}$.

In the effective mass, dielectric constant approximation, the quantum well 
Hamiltonian $H$ can be quite generally written as $H=H_{KS} + V_{res}$, 
where $H_{KS}$ is the Kohn-Sham (KS) one-body Hamiltonian consisting of the 
kinetic, Rashba, Dresselhaus, 
exchange-correlation KS potential and Zeeman terms, and $V_{res}$ is 
the residual Coulomb interaction. The $KS$ Hamiltonian reads

\begin{eqnarray} 
H_{KS} = \sum_{j=1}^N\left[\frac{P^+P^- + P^-P^+}{4m}+ 
\frac{\lambda_R}{2 i\hbar}(P^+\sigma_- - 
P^-\sigma_+) +\frac{\lambda_D}{2\hbar}(P^+\sigma_+ + P^-\sigma_-)\right. 
\nonumber\\ 
\left. +W_{xc}(n,\xi,{\cal
V})\sigma_z + \frac{1}{2}g^* \mu_B B \sigma_z\right]_j\; ,~~~~~~~~~~~~~ 
\label{eq3}
\end{eqnarray} 
where $m=m^* m_e$ is 
the effective electron mass in units of the bare electron mass $m_e$, 
$P^{\pm}=P_x \pm i P_y$, and 
$\sigma_{\pm}=\sigma_x \pm i\sigma_y$. Although other approaches may be also 
considered, we have considered the exchange-correlation potential
$W_{xc}(n,\xi,{\cal V})$
in the local-spin current density approximation 
(LSCDA).\cite{Fer94,Lip03} It depends on the density $n$,
magnetization $\xi=n^{\uparrow}- n^{\downarrow}$, and local 
vorticity ${\cal V}$, and is evaluated from the exchange-correlation energy
per electron ${\cal E}_{xc}$ as $W_{xc}= \partial(n {\cal E}_{xc})/
\partial \xi$. The last term in Eq. (\ref{eq3}) is the
Zeeman energy, where
$\mu_B = \hbar e/(2 m_e c)$ is the Bohr magneton, and $g^*$ is the effective 
gyromagnetic factor. For bulk GaAs, 
$g^*=-0.44$, $m^*=0.067$, and the dielectric constant is $\epsilon=12.4$.
To simplify the expressions, in the following we
shall use effective atomic units 
$\hbar=e^2/\epsilon=m=1$. 

In the following, the residual Coulomb interaction will be treated in the adiabatic 
time-dependent LSCDA (TDLSCDA).\cite{Lip03} We are going to see that,
in the absence of SO coupling, not only the exact Hamiltonian,
but also the one in which the 
residual interaction is treated in the TDLSCDA
fulfill the equation 
\begin{equation} 
\label{eq4}
[H,S_{\mp}]=\pm\omega_L S_{\mp}~, 
\end{equation} 
where 
$S_{\mp}=1/2\sum_j \sigma^j_{\mp}$ and $\omega_L=|g^* \mu_B B|$. Thus, if $|0\rangle$ 
is the gs of the system, the 
states $S_{\mp}|0\rangle$ are eigenstates of $H$ with excitation energies $\pm\omega_L$. 
Note that a negative $g^*$ 
implies that the spin-up states are lower in energy than the spin-down ones, 
and that the actual physical solution of 
Eq. (\ref{eq4}) is that corresponding to the $S_-$ operator. This is the physical
contents of the Larmor theorem. Note also that in the absence of spin-orbit coupling, 
$[H, \sum^N_j P^+_j]=\omega_c \sum^N_j P^+_j$, where 
$\omega_c=eB/(mc)$ is the cyclotron frequency. This is the Kohn theorem, 
which also holds in the adiabatic time-dependent 
local spin density approximation (TDLSDA) and in the TDLSCDA,
and can be generalized to the case of quantum wires and dots parabolically confined.

Since
\begin{equation} 
\label{eq5}
[H,S_-]=
\omega_L S_- +4\sum_{j=1}^N[\lambda_D P^+ \sigma_z+i\lambda_R P^-\sigma_z]_j~,
\end{equation} 
the spin-orbit terms in Eq. (\ref{eq3}) mix the transverse spin excitations
induced by the operator $S_-$ with the 
spin-density excitations induced by $\sum_{j=1}^N P^{\pm}_j\sigma_z^j$,
and thus Larmor's theorem is not fulfilled. In the following, we use 
the equation of motion approach to find the eigenvalues and
eigenstates of the KS Hamiltonian $H_{KS}$ Eq. (\ref{eq3}) which arise from the
SO mixing, and will evaluate the
spin wave dispersion relation $\omega(q)$ by taking into account the effect of 
the residual interaction. This is done by firstly solving  the equation of motion
\begin{equation} 
\label{eq6}
[H_{KS},O^+]=\omega O^+~,
\end{equation} 
and 
then calculate the transverse response
$\chi_t(q,\omega)$ per unit surface ${\cal A}$ in the TDLSCDA:
\begin{equation} 
\label{eq7}
\chi_t(q,\omega)={\chi_t^{KS}(q,\omega)\over1-2F_{xc}\chi_t^{KS}(q,\omega)} \; , 
\end{equation} 
where $F_{xc}=W_{xc}/\xi$, and $\chi_t^{KS}(q,\omega)$ is the KS transverse
response per unit surface.\cite{Lip03} The poles of
$\chi_t(q,\omega)$ yield $\omega(q)$. The transverse spin response without 
inclusion of spin-orbit coupling has been 
studied in the past in the RPA\cite{Kal84} and time-dependent Hartree-Fock\cite{Mac85}
approximations. 

Up to second order in $\lambda_{R,D}$, Eq. (\ref{eq6}) is straighforwardly solved by
the operator 
$O^+=\sum_{j=1}^N [a \sigma_- + b P^+\sigma_z + c P^-\sigma_z +d\sigma_+]_j$.
To do so, one has to use the commutators 
$[\sigma_+,\sigma_-]=4\sigma_z$ and $[P^-,P^+]=2\omega_c$. This yields a homogeneous 
system of linear equations for the 
coefficients $a$, $b$, $c$ and $d$ from which the energies $\omega$ are obtained by 
solving the secular equation valid 
up to $\lambda^2_{R,D}$ order 
\begin{equation} 
\label{eq8}
(\omega^2-\tilde\omega_L^2)(\omega^2-\omega_c^2)
-4\,\omega_c(\lambda_D^2+\lambda_R^2)\omega^2 -
4\,\omega_c^2\,\tilde\omega_L(\lambda_D^2-\lambda_R^2)=0 ,
\end{equation} 
where $\tilde\omega_L \equiv |g^* \mu_B B+2W_{xc}|$.
This quartic equation can be exactly solved, yielding the excitation energies
(only positive solutions are physical). For 
each of them, the homogenous linear system, supplemented with the
normalization condition 
$\langle0\vert[(O^+)^{\dagger},O^+]\vert0\rangle=1$, determines the coefficients
$a$, $b$, $c$ and $d$. We have found it
more convenient to discuss the solutions of the above equation in the
limits
of small and large magnetic fields, which are more transparent and easier to 
compare with available experimental data. In the small $B$ limit 
$(\tilde{\omega}_L, \omega_c \ll  \lambda_R, \lambda_D)$
we obtain  a unique solution
\begin{equation} 
\label{eq9}
\omega= 2 \,\sqrt{\omega_c(\lambda_R^2+\lambda_D^2)} \; .
\end{equation} 
In the large $B$ limit we obtain
\begin{equation} 
\label{eq11}
\omega(S_{\mp})=\pm\left(\tilde\omega_L+2\lambda_R^2{\omega_c\over\omega_c+\tilde\omega_L} 
-2\lambda_D^2{\omega_c\over\omega_c-\tilde\omega_L}\right) \; ,
\end{equation} 
which are mainly excited by the operators 
$S_-$ and $S_+$,  and 
\begin{equation} 
\label{eq12}
\omega(P^{\pm}\sigma_z)=\pm\left(\omega_c-2\lambda_R^2{\omega_c\over\omega_c+\tilde\omega_L} 
+2\lambda_D^2{\omega_c\over\omega_c-\tilde\omega_L}\right) \; ,
\end{equation} 
which are mainly excited by the operators 
$\sum_{j=1}^N [P^+\sigma_z]_j$ and $\sum_{j=1}^N [P^-\sigma_z]_j$. 
By mainly we mean that the coefficient 
of the corresponding operator entering the definition of  is 
$O(\lambda^0_{R,D})$, whereas
all the others are $O(\lambda^2_{R,D})$. Note that if $\lambda_{R,D}=0$,
the two physical modes in the preceding equations are uncoupled.

Equation (\ref{eq9}) shows that,
at $B\sim0$, to order $\lambda^2_{R,D}$ there is no
spin splitting due to the SO coupling.  
Indeed, when $B\to0$ not only $\omega_L$ and $\omega_c$ vary linearly 
with $B$, but also  $W_{xc}$ does, implying that the solution of
Eq. (\ref{eq8}) goes to zero in this limit.
Earlier electron-spin resonance measurements on GaAs quantum
wells\cite{Ste82}
seemed to indicate that a finite spin splitting was present in the 
$B=0$ limit.
However, subsequent experiments carried out by the same group\cite{Dob88}
covering a broader $B$ range point out that the spin splitting of a
Landau level is an exact quadratic function of $B$, and that its
extrapolation to $B=0$ leads to a vanishing spin splitting.
Our result, which  is not changed
by the effect of the residual interaction,  is thus in full 
agreement with the experimental findings of Dobers et al.\cite{Dob88}

{\bf We have checked that, at low $B$ fields, the dominant component of
$O^+$ corresponding to the energy Eq. (\ref{eq9})
is the spin-flip operator $\sum_{j=1}^N[\sigma_-]_j$.
In this limit, the Dresselhaus and Rashba
SO interactions act ``in phase'', whereas at high $B$ they 
partially compensate each other [compare the energy given in
Eq. (\ref{eq9}) with these in Eqs. (\ref{eq11}) and (\ref{eq12})].
This arises from the structure of the secular Eq. (\ref{eq8}), where in the 
low $B$ limit, the second term dominates over
the third one, whereas in the high $B$ limit both terms
are equally important, yielding the solutions shown in  Eqs.
(\ref{eq11}) and (\ref{eq12}).
We have not been able to find a deeper explanation
for this different behavior at low and high magnetic fields.
It is worth to mention that the independent particle Hamiltonian can be
exactly solved when only the Rashba or Dresselhaus SO terms are
included.\cite{Sch03} The merit of  Eqs. (\ref{eq9}-\ref{eq11}) is that
they are exact to the relevant $\lambda^2_{R,D}$ order when both SO
couplings are simultaneously taken into account.
}

The excitation energy $\omega(S_-)$ is the independent particle (KS) value 
for the spin splitting and violates Larmor's theorem even if the SO
coupling is neglected. On the contrary, when the residual 
interaction is properly taken into account, the theorem Eq. (\ref{eq4}) is
recovered.
In the following, we will concentrate on the large $B$ limit.
It is then possible to derive the spin wave dispersion relation,
including spin-orbit effects, by solving the equation 

\begin{equation} 
\label{eq13}
1-2F_{xc}\chi_t^{KS}(q,\omega)=0 
\end{equation} 
that gives the poles of the transverse 
response function Eq. (\ref{eq7}). To do so, we write the transverse spin response
as\cite{Lip03} 
\begin{equation} 
\label{eq14}
{\cal A}\,\chi_t^{KS}(q,\omega)= {\vert\langle\omega(S_{-})\vert{1\over2}\sum_{j=1}^N
e^{i\bf{q}\cdot\bf{r}}\sigma^j_{-}\vert0\rangle\vert^2\over\omega-\omega(S_-)} 
-{\vert\langle\omega(S_{+})\vert{1\over2}\sum_{j=1}^N 
e^{i\bf{q}\cdot\bf{r}}\sigma^j_{+}\vert0\rangle\vert^2\over\omega+\omega(S_+)} \; , 
\end{equation} 
where 
$\vert\omega(S_{\mp})\rangle \equiv O^+[\omega(S_{\mp})]\vert0\rangle$, and the
corresponding energies are given by Eqs. (\ref{eq11}) and (\ref{eq12}).
The calculation of the matrix elements in Eq. (\ref{eq14}) must be done with care since
$|0\rangle$ and $|\omega(S_{\pm})\rangle$ are not eigenstates of $S_z$
because of the spin-orbit coupling. 
Neglecting terms in $\lambda^2_{R,D}/\omega_c$ or smaller, one gets\cite{note1}
\begin{equation} 
\label{eq15}
\chi_t^{KS}(q,\omega)=
\xi\,{|F(q)|^2\over\omega-\omega(S_-)} \; ,
\end{equation} 
where  
\begin{equation} 
\label{eq16}
F(q)={1\over N}\langle0\vert\sum_{j=1}^N e^{i\bf{q}\cdot\bf{r}_j}\vert0\rangle  
\end{equation} 
is the gs elastic form factor. Note that $F(0)=1$ and that 
$F(q)$ goes to zero when $q\to\infty$.
From Eq. (\ref{eq13}) one finally obtains

\begin{equation}
\label{eq17}
\omega=|g^*\mu_B B|+2\lambda_R^2{\omega_c\over\omega_c+\tilde\omega_L} 
-2\lambda_D^2{\omega_c\over\omega_c-\tilde\omega_L} 
-2W_{xc}\left(1-|F(q)|^2\right)
\end{equation}
This is the main result of our work, together with the  
lack of SO splitting we have found in the small $B$ limit.
In the limit $q\to\infty$, Eq. (\ref{eq17}) yields
the independent particle spin splitting Eq. (\ref{eq11}), which
crucially depends on the actual value of $W_{xc}$ entering the
definition of $\tilde\omega_L$.
In the $q=0$ limit, neglecting terms of order
$\tilde\omega_L/\omega_c$, Eq. (\ref{eq17}) reduces to\cite{note1}

\begin{equation}
\label{eq18}
\omega=|g^*\mu_B B|+2(\lambda_R^2-\lambda_D^2) \; .
\end{equation}
This expression shows that, neglecting the SO coupling, 
Larmor's theorem is fulfilled in the adiabatic 
TDLSCDA (it can be shown that the same holds in the adiabatic TDLSDA),
and that at high magnetic fields, the SO 
interaction yields a $B$ independent contribution to the spin splitting.
Taking e.g. $m\lambda_R^2/\hbar^2 = 27$ $\mu$eV,
$m\lambda_D^2/\hbar^2 = 6$ $\mu$eV, which have been recently used to
reproduce the spin
splitting in quantum dots\cite{Kon05} and the splitting of the cyclotron 
resonance in quantum wells,\cite{Ton04} we get
$2m(\lambda_R^2-\lambda_D^2)/\hbar^2\sim40$ $\mu$eV. This is
definitely a small amount, but it may have an influence on
the fine analysis of some experimental results
(note the vertical scale in Fig. \ref{fig1}).

\section{Comparison with experiments and discussion}

\begin{figure}[t] 
\centerline{\includegraphics[width=16cm,clip]{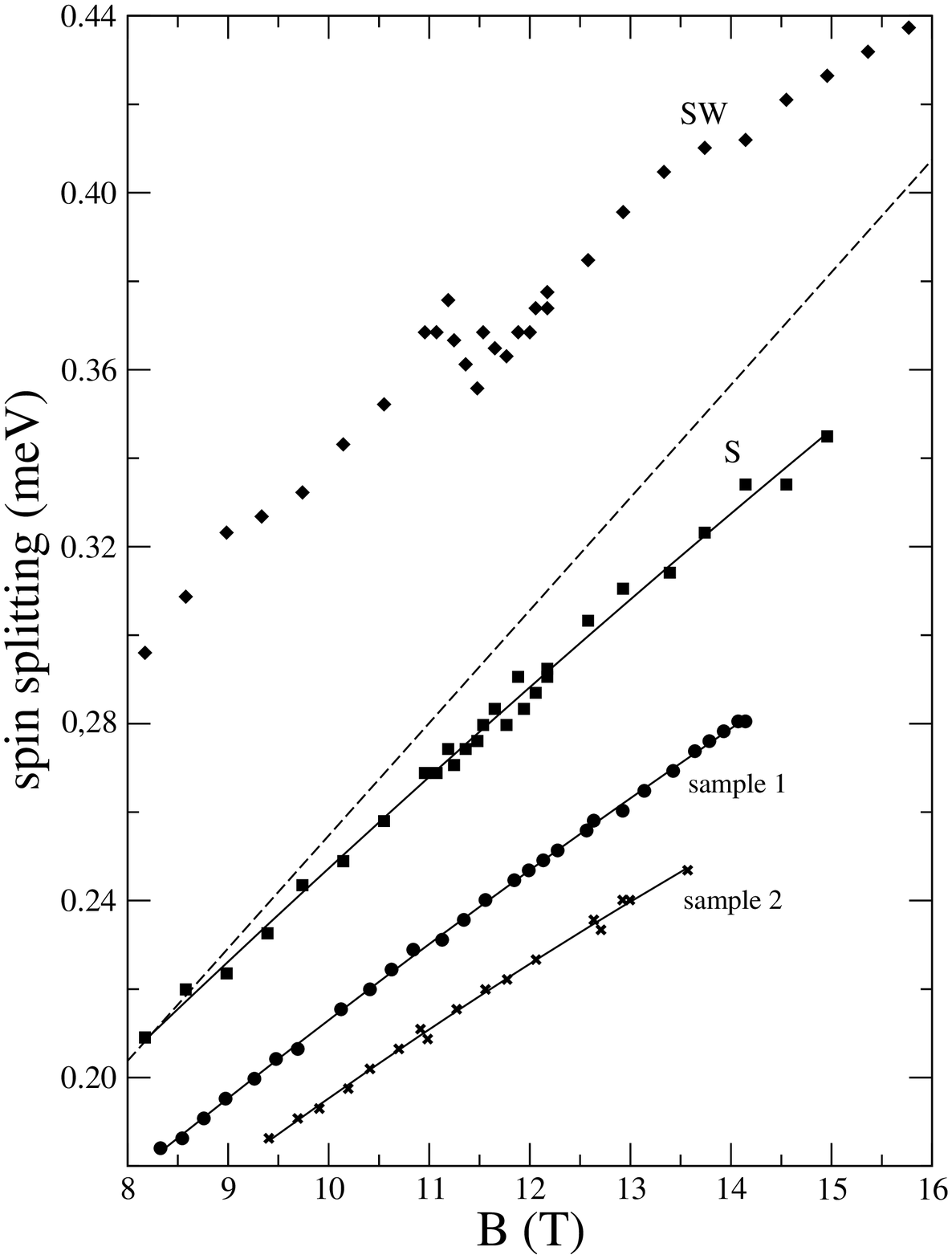} }
\caption{ 
Experimental spin splittings as a function of $B$.
Squares and diamonds, from Ref.~\onlinecite{Dav97}; dots (sample 1)
and crosses (sample 2),
from Ref.~\onlinecite{Dob88}. The dashed line corresponds to $\mu_B g^* B$
with $|g^*|= 0.44$, and the solid lines are fits using the law Eq. (\ref{eq18}).
} 
\label{fig1} 
\end{figure}

Using inelastic light scattering, Davies et al.\cite{Dav97} and Kang 
et al.\cite{Kan00}   have measured charge
and spin density excitations in 2D electrons systems confined in 
GaAs quantum wells at high $B$. In the following we only discuss the 
results of Ref.~\onlinecite{Dav97} because the information presented
in Fig. 2 of this reference is especially well suited for the purpose of
our work.
These results are represented in Fig. \ref{fig1}. The data labeled $S$ 
correspond to wave-vector allowed scattering from the $q=0$ Larmor mode; 
indeed, the maximum in-plane $q$ allowed by the experimental geometry is 
small, $q_{max}= 6 \times 10^4$ cm$^{-1}$, so that the $S$ mode of energy 
$\epsilon_S$ should correspond to the spin splitting energy Eq. (\ref{eq18}).
The data labeled $SW$ is attributed to disorder-activated
scattering, and would correspond to $q\ne0$ excitations of energy 
$\epsilon_{SW}$.\cite{Dav97} The difference between 
$\epsilon_{SW}$ and $\epsilon_{S}$ is attributed
to the exchange enhancement of $\epsilon_{SW}$ above the Zeeman 
energy.\cite{Dav97}
The dashed straight line represents the Larmor energy taking for $g^*$ the bulk 
value, $|g^*|=0.44$. This overestimates $\epsilon_S$, especially at high $B$.

Any sensible comparison with these results must take into account the
$B$ dependence of $g^*$. This dependence has been clearly established in
magnetoresistivity experiments,\cite{Dob88}
taking advantage of the fact that the
electric-spin resonance affects the magnetoresistivity of the 
2D electron gas, and this can be used to determine $g^*(B)$.
These experiments probe the one-electron energy levels and are not
influenced by many-electron interactions, contrarily to magnetoquantum
oscillations, which are strongly influenced by many-electron interactions.

The spin splitting obtained in Ref.~\onlinecite{Dob88} is
represented in Fig. \ref{fig1} as dots and crosses,
which correspond to two different samples.
In the lowest Landau level, which is the common physical situation
for all data represented in Fig. \ref{fig1}, these authors have fitted
$g^*(B)$ as $|g^*(B)|=g_0^* + r B/2$. The value of
the parameters $r$ and $g^*_0$ turns out to sensibly depend
on the experimental sample, and
the possibility of a SO shift at high $B$ values
could not be considered there. Moreover,
the $\lambda_{R,D}$ values are rather poorly known and
dependent on, e.g., the thickness of the experimental sample.

We have thus renounced to use the laws $g^*(B)$ obtained in
Ref.~\onlinecite{Dob88} in conjunction with Eq. (\ref{eq18}),
to establish a clear evidence of  spin-orbit effects on the 
$\epsilon_S$ energy obtained from
resonant inelastic light scattering experiments, and have
satisfied ourselves with the more limited scope of
using Eq. (\ref{eq18})
as a three-parameter law to fit $\epsilon_S$ as well as the spin
splittings of Ref. \onlinecite{Dob88}, with the aim of seeing
whether a reasonable value for these parameters can be extracted.

The solid straight lines in Fig. \ref{fig1} represent the result of
such linear fits, whose parameters are collected in
Table \ref{table1}.
In the case of inelastic light scattering, the neglect of
the SO term in Eq. (\ref{eq18}) yields an
unrealistic $g^*_0=0.49$, as this value should be smaller
than that of bulk GaAs due to the penetration
of the electron wave functions into the Al$_x$Ga$_{1-x}$As
barriers.
The dispersion of the electron-spin resonance datapoints\cite{Dob88}
seems to be smaller, and the analysis of the high $B$ data might
be used to ascertain which SO mechanism is dominating in
a given sample. This could be an alternative {\bf or complementary} 
method to the recently proposed\cite{Kon05} of using the anisotropy of
the spin splitting in single-electron resonant tunneling
spectroscopy in lateral quantum dots submitted to
perpendicular or parallel magnetic fields.
The analysis of samples 1 and 2 would indicate that in the
former, the Dresselhaus SO is the dominating mechanism, whereas
in the later it is the Bychkov-Rashba one.
{\bf We want to stress that
we have extracted the experimental data from a careful digitalization
of the original figures. Due to the smallness of the effects we are discussing, we
cannot discard that this procedure may have had some effect on the value of the
parameters determined from the fit, and our analysis should be considered as
qualitative to some extent. However, we find it encouraging that the 
parameters obtained from the fits are meaningful, and within the range of
values found in other works.\cite{Kon05,Ton04}
}

We finally discuss briefly the $q\ne0$ $SW$ mode. From Eqs. (\ref{eq17})
and (\ref{eq18}), we have that
$\epsilon_{SW}-\epsilon_S=-2W_{xc}(1-|F(q)|^2)$.
At high $q$, this difference is sensibly determined by $W_{xc}$.
In our calculation, as well as in time-dependent Hartree-Fock\cite{Mac85} and
exact diagonalization\cite{Rez87} calculations, we have found
values of $W_{xc}$ of the order of $-2$ meV. Hence, $-2W_{xc}$ is about a factor
40 larger than the measured $\epsilon_{SW}-\epsilon_S$, which has been 
obtained in the $q\to0$ limit where short range correlations
are very important in determining the actual value of $1-|F(q)|^2$.
This can be seen by assuming for $F(q)$ the independent particle value.
Using a Slater
determinant made of Fock-Darwin single particle wave functions to describe the gs
of the system at $B\neq0$, one finds $F(q\ell)=e^{-q^2\ell^2/4}$,
where $\ell$ is the magnetic length $\ell=(\hbar c/e B)^{1/2}$.
In the small $q$ limit, $\epsilon_{SW}-\epsilon_S\simeq-W_{xc}q^2\ell^2$. 
For $B=10$ T, $q_{max}\ell\simeq 0.05$ and
$\epsilon_{SW}-\epsilon_S \sim 0.01$ meV,  which is about one tenth
of the experimental result as shown in Fig. \ref{fig1}. 
Light scattering experiments at small $q$ and high $B$ are thus
very sensitive to correlation
effects in the elastic form factor, which is the key quantity to reproduce
the experimental findings.  

\section*{ACKNOWLEDGMENTS} This work has been performed 
under grants FIS2005-01414 from 
DGI (Spain) and 2005SGR00343 from Generalitat de Catalunya.
E. L. has been suported by DGU (Spain), grant SAB2004-0091.

\pagebreak

\newpage

\begin{table}
\caption{
Parameters of the least-square linear fit to the cited experimental data.
}

\begin{tabular}{|c|c|c|c|} \hline
           & $\hbar^2 m(\lambda_R^2-\lambda_D^2)$  & $g^*_0$ & $r$         \\
           & ($\mu$eV) &    & (T$^{-1}$) \\ \hline
$\epsilon_S$ (Ref. \onlinecite{Dav97})    &  $8.6$     & $\;0.43\;$ & $\; -7.3 \times 10^{-3}\;$ \\
sample1 (Ref. \onlinecite{Dob88})    &   $7.7$     & 0.38    & $ -8.1 \times 10^{-3}$ \\
sample2 (Ref. \onlinecite{Dob88})   &   $-1.3$     & 0.41   & $ -1.3  \times 10^{-2}$ \\
\hline \end{tabular}
\label{table1}
\end{table}


\begin{thebibliography}{99}
\bibitem{note1} We have checked that the results presented in Sec. III
remain unchanged if these terms are kept in the calculation.
\end{thebibliography}
\end{document}